\documentclass[aps,prl,twocolumn,unsortedaddress,floatfix,longbibliography]{revtex4-2}

\usepackage{tabularx}
\usepackage{parskip}
\usepackage{amsthm}
\usepackage{amsfonts}
\usepackage{siunitx}
\usepackage{amsmath}
\usepackage{amssymb}
\usepackage{graphicx}
\usepackage{verbatim}
\usepackage[colorlinks=false,colorlinks]{hyperref} 
\usepackage{tikz}
\usepackage{braket}
\usepackage{xcolor}
\usepackage{color}

\begin{document}

\title{ Observation of Magnetic Devil’s Staircase-Like Behavior in Quasiperiodic Qubit Lattices}

\author{Alejandro Lopez-Bezanilla }
\email[]{alejandrolb@gmail.com}
\affiliation{Theoretical Division, Los Alamos National Laboratory, Los Alamos, New Mexico 87545, USA}

\date{\today}

\begin{abstract} 
The devil’s staircase (DS) phenomenon is a fractal response of magnetization to external fields, traditionally observed in periodic ferromagnetic systems, where the commensurability between spin arrangements, lattice parameters, and external magnetic fields governs abrupt changes in magnetization. Its occurrence in aperiodic, fractal-type systems has remained largely unexplored, despite their natural compatibility with such phenomena. 
Using a quantum annealing device, we uncover a wealth of abrupt magnetic transitions between spin manifolds driven by increasing external magnetic fields within a simple yet effective Ising-model framework.
In contrast to periodic systems, where DS arises from long-range competing interactions, our findings reveal that short-range, purely antiferromagnetic couplings in aperiodic geometries produce equally rich ground-state magnetization patterns. We demonstrate that while  magnetic textures are determined by the lattice size, their formation remains remarkably robust and independent of scale, with commensurability emerging locally. 
Our results challenge the prevailing view that DS behavior is limited to periodic systems and establish quasiperiodic geometries as a natural host for this phenomenon.

 \end{abstract}
 
\maketitle

\section{Introduction}
The behavior of magnetic materials is determined by the interaction between their atomic arrangement and spin organization, as well as the multistable processes induced by external stimuli such as magnetic fields or temperature~\cite{Flahaut}. Crystalline systems with modulated commensurate or incommensurate structures are particularly interesting in this regard, as they illustrate how the delicate balance between external forces and interatomic interactions can drive a wide range of phase transitions~\cite{Zapf2018}. The devil’s staircase (DS)~\cite{Bak,Bruinsma,refId0} framework captures the stepwise sequence of transformations in these systems, where each magnetization step corresponds to a commensurate structure with a wave vector locked to a rational fraction $n/m$~\cite{Rossat}. The fractal nature of this hierarchy, with an infinite sequence of intermediate steps, is the material's response to a control parameter and arises from the complex interplay of competing magnetic interactions within the material~\cite{JVillain_1977}, particularly those exhibiting spin-glass behavior or incommensurate magnetic order~\cite{Edwards_1975}.

Previous studies have explored the interplay between commensurability and frustration, as well as the role of dimensionality and disorder in shaping the behavior of DS~\cite{ Baibich88, Binasch89,Matsuda2015}. Exploration of magnetic rare earth compounds~\cite{Elliott61}, quasi-one-dimensional conductors~\cite{PhysRevLett.98.217201}, and thin magnetic films~\cite{Baibich88,Binasch89} has revealed insights into the role of commensurability in dictating the nature of magnetic transitions and shaping the magnetic response.
When the external magnetic field aligns with the inherent periodicities of a ferromagnetic material, commensurate magnetic states emerge,  leading to well-defined steps in the magnetization or susceptibility~\cite{Agrestini08}. These commensurate states are characterized by long-range order and exhibit rational fractional values of the saturation magnetization. In contrast, incommensurate conditions give rise to fractional steps or irregularities in the magnetic response, reflecting the presence of incommensurate spin density waves or soliton-like structures~\cite{Machida}. 

\begin{figure}[htp]
 \includegraphics[width=0.98 \linewidth]{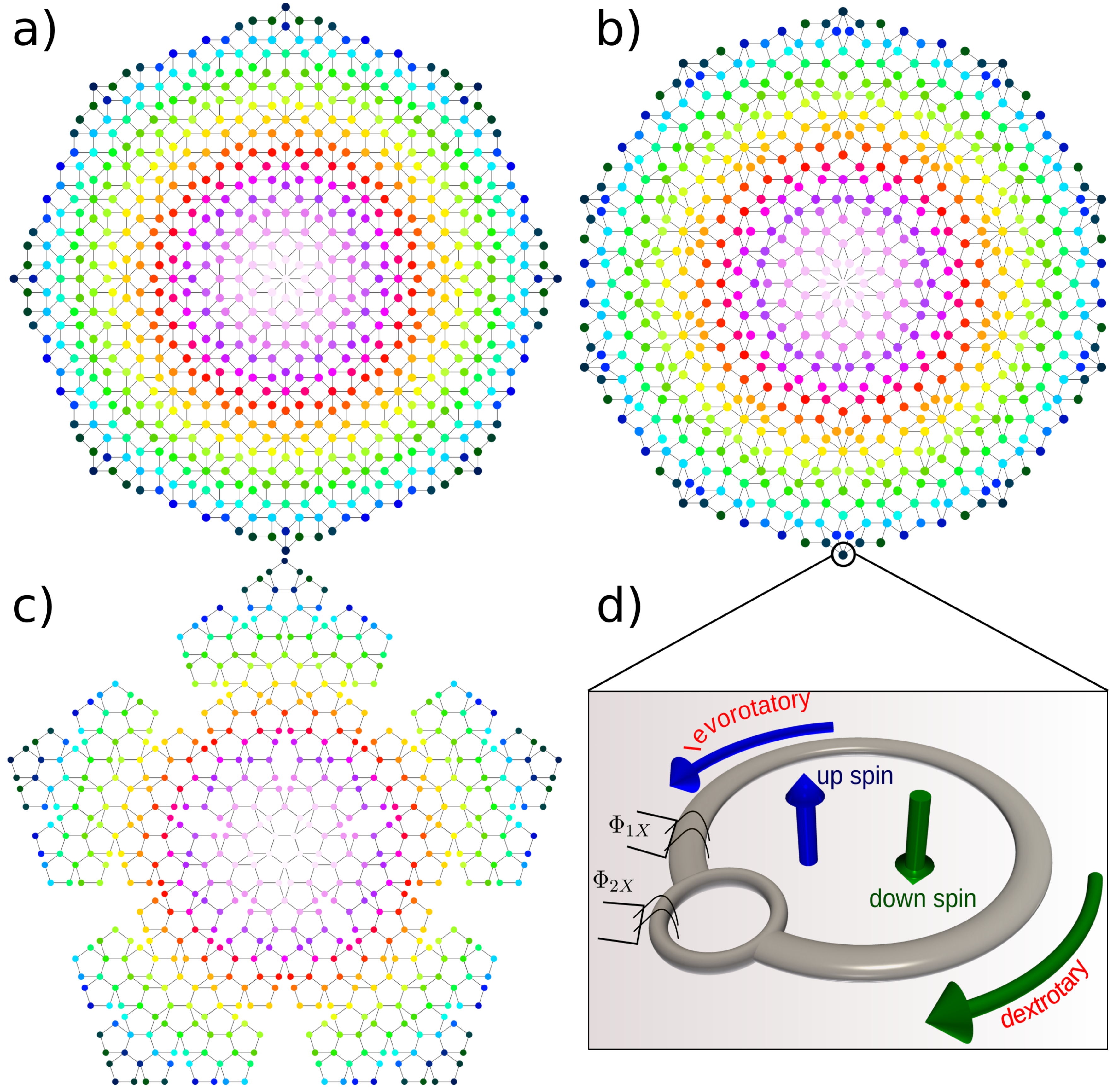}
 \caption{ Graphs depict three quasilattices, with colors representing automorphism groups (see text): 55 (AB17), a), 42 (AB16), b), and  60 (PPL) c).  
 d)  The two basis states, $\ket \uparrow$ and $\ket \downarrow$, correspond to out-of-plane magnetic fields with opposite orientations, generated by persistent currents circulating in opposite directions within the superconducting loop.
 } 
\label{fig1}
\end{figure}

In this paper, we study related phenomena using quasiperiodic structures to explore the interplay between complex geometries, short-range order, and magnetic interactions, offering a different perspective on the DS phenomenon complementary to those observed on ferromagnetic periodic crystalline lattices. 
The aperiodicity and rich local environments of quasiperiodic lattices offer a diverse set of coordination numbers and connectivity patterns, that naturally supports the emergence of DS-like behavior and support the stabilization of multiple, energetically distinct spin configurations under varying external magnetic fields. The uniform local environments of periodic lattices typically yield a limited number of commensurate plateaus, while disordered or amorphous lattices tend to suppress abrupt transitions and instead produce smooth, continuous magnetization responses as a result of irregular couplings and the lack of structural constraints.
Our study focuses on regular, symmetrical, and isotropic topologies, characterized by $n$-fold coordination ($n=2, ... ,10$) of vertices, which enables a diverse array of uniform and well-regulated magnetic phases, exhibiting a monotonic evolution of the average magnetization. This behavior is notably rare in conventional crystals and thin-layered ferromagnetic materials~\cite{Elliott61,Baibich88,Binasch89}.
Quantum simulation is conducted on a quantum annealer (QA)~\cite{KingPRX,Kairys2020, King2022_2000,King2023,Arctic2023}, a novel approach that allows for the design and creation of aperiodic tilings.

\section{Quasiperiodic geometries and materials}

Quasiperiodic patterns from higher-dimensional periodic structures are useful for describing quasicrystalline materials, offering a noncrystallographic counterpart to traditional crystallographic cell models. Quasiperiodic structures exhibit rich interconnectivity of the node network and features arising from their local similarity and lack of translational symmetry, leading to fractal-like geometries. Figure~\ref{fig1} shows two types of aperiodic two-dimensional structures. Firstly, the so-called Ammann-Beenker (AB) tiling~\cite{grunbaum1987tilings} is a two-dimensional octagonal quasilattice composed of squares and 45$^\circ$ and 135$^\circ$ rhombi, featuring a rich junction landscape where intersections of squares and rhombi can be observed in Figure~\ref{fig1}a . A modification in the angles of these prototiles leads to a variation of the structure that expands the connectivity on the final graph, as shown in Figure~\ref{fig1}b in a second version where the central node is now connected to ten other nodes and the squares are now transformed into rhombi.
A third planar shape of interest is the pentaplexity (PPL) tiling Figure~\ref{fig1}c, created from a regular pentagon that is subdivided into six smaller ones,  leaving slim triangular gaps~\cite{Penrose1979}. Although the subdivision process generating each quasiperiodic structure can proceed indefinitely, we restrict our analysis to finite realizations containing up to 705, 641, and 560 nodes, respectively.

Although these geometries are defined as a mathematical construct, they find a correspondence with atomic arrangements in actual materials. Electron diffraction of an eightfold rotational axis quasilattice was obtained from rapidly solidified V-Ni-Si~\cite{Wang} and Mn-Si-Al~\cite{Wangg} alloys. Using nanofabrication techniques~\cite{RevModPhys.85.1473, Collins2017,Bhat2013,Noya2021,Smetana2020}, tenfold rotational symmetry photonic~\cite{Levi2011,Vardeny2013}, magnonic~\cite{Watanabe}, molecular quasiperiodic~\cite{Collins2017,Urgel2016}, and graphene-based~\cite{Ahn2018} structures have been experimentally designed.

These quasilattices can be represented as graphs, wherein nodes are interconnected by edges, and are categorized based on their respective automorphism groups. An automorphism within a graph entails a permutation of its nodes that maintains the adjacency relations, effectively preserving the graph's structure. By identifying the automorphism group associated with each node, including all permutations that retain the node's identity, nodes can be grouped into equivalence classes, indicative of their symmetrical attributes and local contexts. This feature is exploited to selectively adjust subsets of nodes during the QA shimming procedure~\cite{tutorial}, refining the parameters defining the Hamiltonian to counteract potential biases or inaccuracies stemming from qubit and couplers asymmetries. The color scheme on the graphs of Figure~\ref{fig1} codes the automorphism groups with a total number of 55, 42, and 60 for a, b and c, respectively. 

\section{Quantum annealing}

In our study, a D-Wave programmable QA is employed to explore and extract thousands of low-energy spin configurations. Quantum annealing leverages quantum fluctuations to explore the energy landscape efficiently, exploiting quantum tunneling effects to traverse barriers and find deep minima in the solution space, as opposed to mere simulations on classical hardware with classical methods that often struggle with large-scale combinatorial optimization problems and local minima.
It is worth pointing out that the utilization of a QA offers the advantage of observing real physical processes, as it serves as a surrogate platform for directly observing interactions with external fields. This approach bypasses the reliance on heuristic approximations typically necessary in classical simulations, providing results that are   straightforward predictions of material behavior based on robust models.

Quantum simulation was conducted using a superconducting QA processor in which the two lowest-energy eigenstates of a qubit correspond to macroscopic persistent currents circulating in opposite directions in a loop of an inductance radio frequency superconducting quantum interference device (rf-SQUID)~\cite{Mooij,PhysRevB.80.052506}. Figure~\ref{fig1}d shows a schematic representation of a superconducting qubit. These circulating currents produce magnetic dipole fields that point perpendicular to the plane of the loop and are used to define the effective spin states $\ket \uparrow$ and $\ket \downarrow$. The classical potential energy landscape of the qubit is a double well, shaped by the external flux bias $\phi_{1x}$, where each well corresponds to one of the two current states. The barrier between the wells is controlled by $\phi_{2x}$,  which tunes the tunneling amplitude between the states.  In our interconnect-qubit setting, qubits are coupled via programmable couplers that mediate two-qubit interactions. Couplings define the interaction graph and determine the global energy landscape in which the collective dynamics unfold under an external magnetic field $h$.

The over 1200 flux qubits of the Advantage2.4 integrated circuit in a Zephyr topology\cite{boothby2020} allow us to embed up to 705-qubit lattices. Each node of a graph corresponds to one qubit, while graph edges represent the coupling between individual qubits, tunable with programmable on-chip control circuitry. The low-energy state of a classical Ising Hamiltonian is found using quantum adiabatic optimization by smoothly evolving a time-dependent Ising Hamiltonian from a trivial quantum superposition to a classical energy function\cite{PhysRevA.93.052320,King2022_2000}.

\begin{figure}[htp]
 \includegraphics[width=0.99 \linewidth]{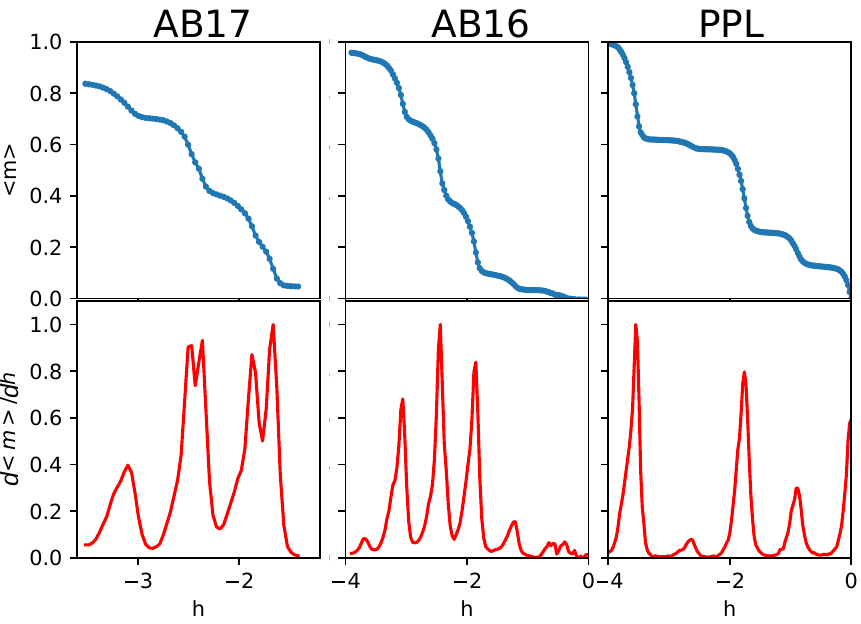}
 \caption{ Average magnetization $\langle m \rangle$ curves of the Ammann-Beenker (AB) geometries shown in Figure~\ref{fig1} (upper panels) and susceptibility, $\chi= d\langle m \rangle /dh$ , (lower panels). } 
\label{fig2}
\end{figure}

The low-energy state of a classical Ising Hamiltonian is found using quantum adiabatic optimization by smoothly evolving a time-dependent Ising Hamiltonian from a trivial quantum superposition to a classical energy function:
\begin{equation}
    \mathcal{H}(s) = - \Gamma(s) \sum_{i}\sigma_{i}^x  + \mathcal{J} (s) \left(  \sum_{i} h_i \sigma_i^z + \sum_{i,j} J_{ij}\sigma_i^z\sigma_j^z \right)
\label{eq}
\end{equation}
where $s=t/t_f$ defined in the $0\textup{--}1$ range is a unitless annealing parameter controlling the annealing progress, such that where $\Gamma(0)\gg \mathcal J(0)$ and $\Gamma(1)\approx 0 \ll \mathcal J(1)$. $\Gamma(s)$ is the transverse field strength that decays to zero for $s = 1$, and $\mathcal{J}$ is an Ising energy scale that increases from zero at $s = 0$. The classical Ising Hamiltonian multiplying $\mathcal{J}$  is determined by one-local longitudinal field $h_i$ and two-local couplings $J_{ij}$. $\Gamma(s)$  multiplies the transverse field Hamiltonian introducing quantum fluctuations in the driver.  

\section{Results and Discussion}

Quantum simulation involving superconducting qubits was carried out to examine the parameters leading to frustrated low-energy arrangements within a nearest-neighbor antiferromagnetic (AFM) spin-$\frac{1}{2}$ Ising model~\cite{Kairys2020,KingPRX}.
The individual-qubit average magnetization $\langle m_i \rangle = \sum^M_{l=1} \sigma^z_{i,l} /M$ is computed by averaging the magnetization at site $i$ over the full ensemble of $M$  collected spin configurations.
In Figure \ref{fig2}, the total magnetization $\langle m \rangle$ of a structure composed of $N$ nodes  is plotted versus the applied magnetic field $h$, where $\langle m \rangle =  \sum_{i=1}^N \langle m_i \rangle /N$. The magnetic susceptibility  is defined as $\chi= d\langle m \rangle /dh$. The upper panels show the monotonic increase in $\langle m \rangle$ with increasing strength of $h$, while the lower panels present the corresponding $\chi$, highlighting geometry-specific patterns in the susceptibility.
Peaks in $\chi$ correspond to critical field values where the system experiences rapid rearrangements in its spin configurations, akin to transitions between different spin manifolds. 

\begin{figure*}[htp]
 \includegraphics[width=0.9 \linewidth]{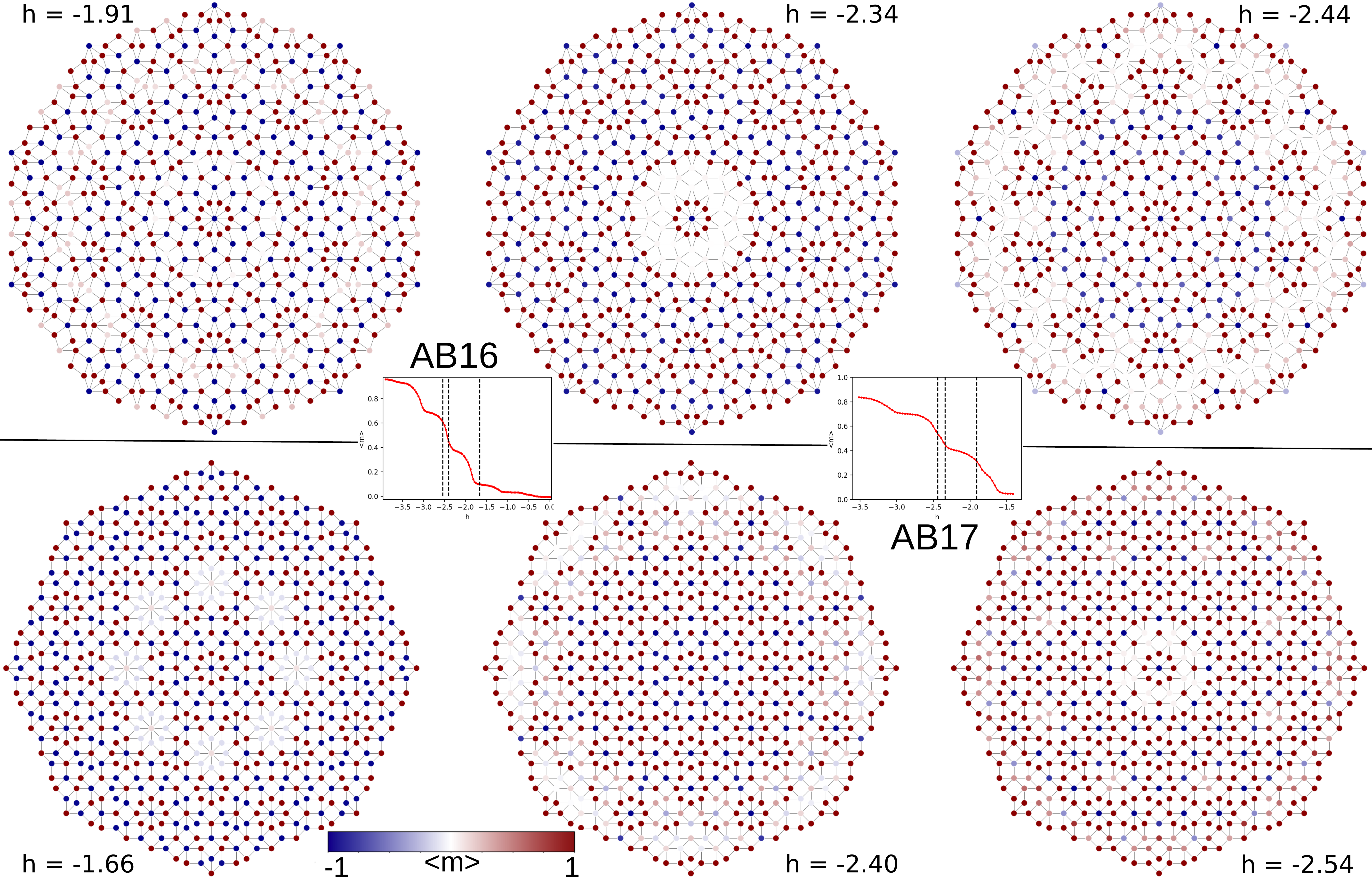}
 \caption{Real space representation of the qubit-resolved $ \langle m \rangle$ in the AB16 (upper panels) and the AB17 (lower panels) quasilattices for selected values of the external magnetic field $h$. In the insets, dashed lines highlight the $h$ values in the $ \langle m \rangle$ plot. The emergent magnetic textures result from the interplay between geometric frustration and the antiferromagnetic interactions with $h$. The distributions reveal the intricate local response as a function of $h$, with distinct patterns of spin alignment and abrupt transitions reflecting the highly degenerate and aperiodic nature of the lattice. The observed configurations are indicative of the underlying symmetry and quasiperiodic order and demonstrate the critical role of external field tuning in driving spin sectors through a series of complex, nontrivial states.   } 
\label{fig3}
\end{figure*}

For each geometry, the magnetization curve exhibits a generally stepwise evolution, with regions of rapid change or smoother plateaus depending on the structure's connectivity and degree of frustration. In all cases, susceptibility peaks accompany transitions in  $\langle m \rangle$, signaling discrete spin reconfigurations as the magnetic field is varied. Differences are visible across geometries, including variations in the sharpness and spacing of these features. The presence of these susceptibility peaks across all geometries supports the interpretation of a DS-like response in both quasiperiodic and periodic lattices.

 
The $\langle m \rangle$ curves reflect structure-wide averages, but individual qubits can exhibit much sharper responses to small variations in the external magnetic field. In certain regions, a minor change in $h$ can induce a complete reversal of a qubit’s spin state and trigger local reconfigurations, while the global magnetization appears to vary less abruptly. Despite this smoothing, the system retains DS-like behavior at the level of its underlying spin configurations. The observed magnetization reflects transitions between discrete, energetically stable spin states that dominate over finite field intervals. 
This behavior is particularly evident, for instance, in the upper central panel of Figure~\ref{fig3}, where the local magnetization in a central region reflects an abrupt evolution across three closely spaced values of the applied field. The figure captures one snapshot of this progression, and reflects the underlying sensitivity of certain qubits to small field variations and highlights the existence of discrete, field-dependent spin configurations characteristic of DS-like behavior: At $h =-2.21$, the spins exhibit a well-defined orientation; for a slightly different $h =-2.34$ (Figure~\ref{fig3}), the same region displays zero local magnetization, consistent with a frustrated  state; and at the next increment, the spins appear fully reversed, supporting the interpretation of magnetization plateaus as arising from transitions between distinct, field-stabilized spin configurations.  The stepwise variation in $\langle m \rangle$ thus emerges from statistical effects acting on spatially distinct regions of the structures.

This qubit-resolved analysis of $ \langle m_i \rangle $ is extended to more regions in Figure~\ref{fig3}, which displays six real-space magnetization maps for the AB16 and AB17 structures, each corresponding to a specific magnetic field value (as indicated in the insets). The panels provide insight into the local magnetic response at each node,  illustrating how individual qubits contribute to the system’s overall magnetization dynamics at the fields indicated in the insets.  These visualizations reveal how magnetization evolves across the lattice upon application of $h$ that enables transitions between distinct spin manifolds, helping to identify localized regions where abrupt changes occur as the field is varied.

These spin sectors can be thought of as discrete domains of magnetization configurations that are locally stable under certain external and geometrical conditions. The transitions between these sectors are driven by the competing interactions—specifically, the short-range AFM couplings, the geometric constraints of the quasiperiodic lattice, and the external magnetic field. As $h$ varies, it selectively lifts degeneracies in the system, destabilizing specific spin configurations, and allowing the system to overcome energy barriers and reorganize into new local minima. Reorganizations of the spin system translate into stable plateaus in the magnetization curve marking distinct sectors and abrupt jumps highlighting critical thresholds. The subsequent jumps in $\langle m \rangle$ reflect the sensitivity of the energy landscape to $h$, which acts as a control parameter guiding the system through a sequence of stable and unstable states.

\begin{figure*}[htp]
 \includegraphics[width=0.99 \linewidth]{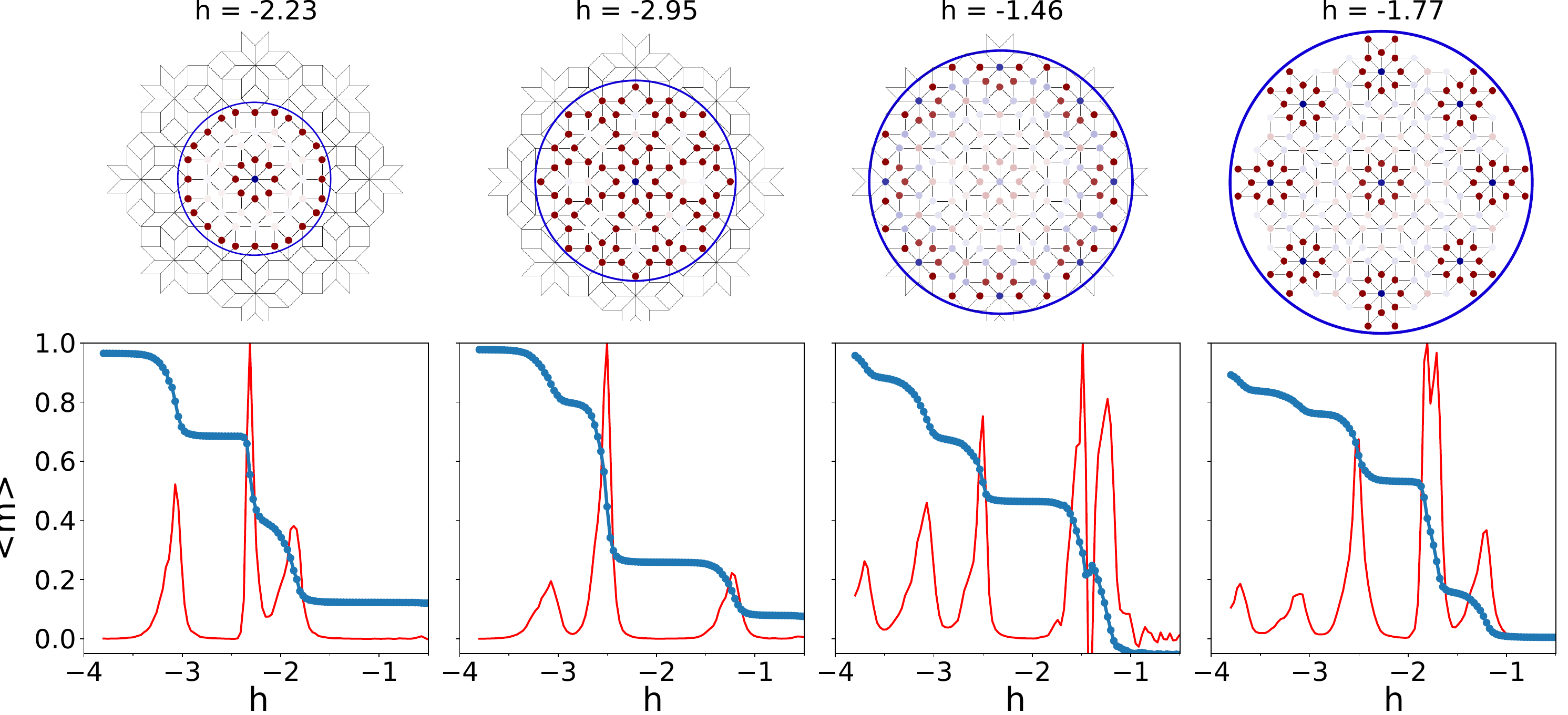}
 \caption{ 
 Lower panels: Evolution of the average magnetization  $\langle m \rangle$ (blue curves) and corresponding normalized susceptibility, $\chi$, (red curves) in the AB17 lattice as a function of system size.  Each panel displays a progressively larger portion of the lattice under a different applied field $h$. Color map of nodes encodes local magnetization (see Figure~\ref{fig3}). Upper panels illustrate the increasing lattice size, with the current finite structure highlighted by a circle. While the specific spin patterns vary, the emergence of organized domains and stepwise magnetization persists, illustrating the robustness of the DS-like response in the quasiperiodic geometry. 
 } 
 \label{fig4}
\end{figure*}

The white-colored regions in the figure, where  $\langle m \rangle \approx 0$ indicate a highly degenerate state where spins are equally likely to adopt one orientation or its opposite. In such regions, the external magnetic field $h$ does not provide a sufficient bias to favor one orientation over the other. The system is left in a state of balance between competing configurations and reflecting a critical feature of frustrated systems: the inability to settle into a unique ground state due to the constraints imposed by local and global interactions. The aperiodic nature of the lattice contributes to this behavior, as the lack of translational symmetry disrupts the regular energy minimization pathways seen in periodic systems\cite{Zapf2018}. As a result, certain regions of the field parameter space correspond to sensitive boundaries between energetically favorable spin configurations, where slight variations in external conditions can lead to transitions between statistically distinct outcomes in the sampling.
These regions of near-zero magnetization could be viewed as boundaries between distinct manifolds, where the system’s spins are delicately balanced and poised to reorganize.

The use of the term ``DS-like behavior'' is justified in this context, despite the distinct commensurability mechanisms at play compared to classical DS systems.
In our AFM lattices, the external field plays an essential role in rebalancing the energy landscape, effectively stabilizing certain spin configurations. While the commensurability here does not strictly arise from intrinsic structural locking, as in traditional DS systems~\cite{Bak,Bruinsma,refId0}, it reflects an external-field-induced alignment of the energy minima with discrete spin configurations.  
While in conventional DS systems plateaus correspond to simple rational fractions of the saturation magnetization, the plateaus observed here do not exhibit an obvious correspondence to such fractional values. Instead of arising from a commensurability condition with a fixed underlying periodicity, the formation of plateaus is governed by the stabilization of discrete spin configurations shaped by the quasiperiodic finite geometry, frustration, and boundary effects. These configurations define spin domains with fixed magnetization values, which vary nontrivially with system size and geometry. 
Decision points in the magnetization landscape mimic the hallmarks of DS-like behavior, and while the absence of rational fractions distinguishes this behavior from classical DS models, the stepwise structure, discrete transitions, and sensitivity to control parameters support the characterization as DS-like. This distinction highlights the combined influence of the two-dimensional frustrated lattice, anisotropic interactions, and external field in driving a modified staircase behavior.

Furthermore, the finite size imposes a natural limit on the number of abrupt steps observable in the magnetization curve. This contrasts with the idealized fractal structure of a classical DS, which assumes an infinite hierarchy of steps. However, the key characteristic of DS-like behavior—self-similarity—is still present, as evidenced by the emergence of new steps at different locations when the system size is increased. 
Figure~\ref{fig4} shows the evolution of $\langle m \rangle$ in the AB lattice as its size increases, for progressively larger lattice portions, highlighting that the characteristic structure of these patterns is preserved and that their formation occurs at any scale. This result suggests that the underlying physics governing the interplay between lattice geometry, frustration, and external magnetic field $h$ remains consistent regardless of the lattice size, even though magnetic domains are different for a same field on two different lattice extensions. Larger lattices introduce greater complexity and variability in the local environments, which is reflected in subtle differences in the magnetization textures. However, the emergence of fractal-like, self-similar structures persists, validating the robustness of the finite quasiperiodic geometry in shaping the magnetic response.
This suggests that the finite system captures a partial representation of the full fractal hierarchy, with its resolution being modified  as the size grows and the lattice boundary changes. Thus, while fractal behavior is always limited in finite systems, its fundamental mechanisms remain intact, linking the finite-size observations to the infinite DS framework.

\section{Conclusion}

To summarize, our study demonstrates how external field-driven transitions in a frustrated antiferromagnet give rise to DS-like behavior, where distinct spin sectors emerge as discrete magnetization plateaus. By leveraging a QA platform, we provide a realization of these frustration-induced transitions in aperiodic lattices showing that while commensurability governs magnetization plateaus in classical periodic systems, the quasiperiodic geometry introduces local commensurability conditions that stabilize similar plateaus. This suggests the DS phenomenon is not inherently tied to long-range interactions or incommensurability in periodic lattices but instead arises from the interplay of frustration, external fields, and local spin configurations, offering a connection to spin ice systems\cite{RevModPhys.91.041003}, where these factors similarly govern phase transitions. The discrete nature of the spin sector transitions observed here echoes the monopole-like excitations in spin ices\cite{Castelnovo2008}, suggesting that similar controlled external stimuli could be used to engineer frustration-driven ordering in artificial spin systems. 
Our results could have potential technological applications ranging from sensing to memory storage, particularly in designing magnetic materials that exhibit a diverse array of magnetic textures tunable with magnetic field. The evidence provided serves as a benchmark for future material realizations, from atomically thin systems to molecular frameworks, and offers a foundation for designing magnetic memory elements where distinct magnetic phases can encode information at the fundamental level.

\section{Acknowledgements}
Work was carried out under the auspices of the U.S. DOE through the Los Alamos National Laboratory, operated by Triad National Security, LLC (Contract No. 892333218NCA000001).
The research presented in this article was supported by the Los Alamos National Laboratory Laboratory (LANL) Laboratory Directed Research and Development (LDRD) program under Project No. LDRD-ER 20240149ER.
I thank Dr. A. M. N. Niklasson for providing the funding support that made this research possible.

\bibliographystyle{naturemag}  
 

\end{document}